\documentclass[twocolumn,showpacs,preprintnumbers,amsmath,amssymb,aps,10pt,prb]{revtex4-1}
\usepackage{graphicx}
\usepackage{verbatim} 
\usepackage[dvips]{color} 

\begin{document}

\title{All-optical implementation of a dynamic-decoupling protocol for hole spins in (In,Ga)As quantum dots}

\author{S. Varwig$^1$}
\author{E. Evers$^1$}
\author{A. Greilich$^1$}
\author{D.~R. Yakovlev$^{1,2}$}
\author{D. Reuter$^{3}$}
\email[Present address: ]{Universit\"at Paderborn, Department
Physik, 33098 Paderborn, Germany}
\author{A.~D. Wieck$^3$}
\author{M. Bayer$^{1,2}$}

\affiliation{$^1$ Experimentelle Physik 2, Technische Universit\"at
Dortmund, 44221 Dortmund, Germany}

\affiliation{$^2$ Ioffe Physical-Technical Institute, Russian Academy of Sciences, 194021
St. Petersburg, Russia}

\affiliation{$^3$ Angewandte Festk\"orperphysik, Ruhr-Universit\"at Bochum, 44780 Bochum,
Germany}

\begin{abstract}
We demonstrate the potential of a periodic laser-pulse protocol for dynamic decoupling of hole spins in (In,Ga)As quantum dots from surrounding
baths. When doubling the repetition rate of inversion laser pulses between two
reference pulses for orienting the spins, we find that the spin
coherence time is increased by a factor of two.
\end{abstract}


\maketitle

Solid state implementations of quantum information technologies have been considered attractive
because of their potential advantages such as robustness, miniaturization, scalability and connection to
conventional information processing
hardware.\cite{Lloyd1993,Burkard2000} The huge obstacle in such
approaches is the 'rigid' coupling of the quantum bits to their
surrounding, strongly limiting their coherence. The demand for
long-lived coherence has quickly geared activities towards carrier
spins in crystals.\cite{Henneberger2008,Ladd2010} For them, carrier
localization suppresses decoherence mechanisms involving orbital
motion.\cite{Dyakonov2008} This has led to the suggestion of
the hyperfine interaction with the surrounding nuclear spin bath as
the main decoherence mechanism for carrier
spins.\cite{Khaetskii2002,Merkulov2002,Luke2009}

For further improvement in this respect, two different strategies can
be pursued: either purification towards zero nuclear spin isotopes
or implementation of protocols for dynamic decoupling from baths.
Also a combination of both strategies may be applied, which has lead, for
example, to an extension of the spin-coherence time associated with the
NV$^-$ center in diamond to milliseconds, even at room
temperature.\cite{Bala2009,Bar2013} For III-V semiconductors
isotope purification is not possible, but dynamic decoupling has shown
great promise. For gate-defined quantum dots the coherence time
could be extended at cryogenic temperatures from a few up to
200\,$\mu$s in that way.\cite{Bluhm2011}

Recently, considerable efforts have been made to develop optimized
pulse sequences reaching a complexity far beyond the initially applied periodic
Carr-Purcell-Meiboom-Gill (CPMG) protocol.\cite{CPMG1958,Uhrig2007,Witzel2007,Lee2008,Jiangfeng2009,Souza2012}
This possibility is offered by the accurate electronic control of
radiation pulses in the microwave frequency range. However, these pulses are limited to durations of more than a
nanosecond. Much shorter pulses are possible employing lasers, but
the possibility to vary the pulse properties within a manipulation
sequence is limited. Due to these difficulties an extension of the spin-coherence time by dynamical decoupling
with laser pulses has not yet been accomplished to the best of our
knowledge, and this is the goal that we target here. 
Dynamic decoupling leads to an insensitivity to noise sources characterized by
coupling frequencies lower than the pulse rate in the implemented protocol, opening up a frequency gap in the interaction with surrounding baths.\cite{Viola,Khodjasteh} Therefore, implementations using pulses as short as possible 
are appealing because of the higher possible pulse rates that may be applied. Under these circumstances the spins become less sensitive to noise sources with enhanced coupling frequencies.

For that purpose we monitor the precession of hole spins confined in
an (In,Ga)As/GaAs quantum dot ensemble about an external magnetic field
after their orientation normal to that field by pump
pulses.\cite{Varwig2012} To study the spin coherence, we optically
stimulate spin echoes by applying rotation pulses, which invert the
inhomogeneous dephasing of the hole spin ensemble precession. 
The underlying basic operation of a single $\pi$-rotation of spins was demonstrated earlier.\cite{DeGreve2011,Varwig2013}
From the echo amplitude dependence on the time after spin initialization the hole
spin coherence time can be accurately assessed.
The echo-inducing pulses are applied periodically, mimicking 
the CPMG protocol used for extending spin coherence in dynamical
decoupling.\cite{footnote} By doubling the repetition rate of rotation pulses we find that
the hole spin coherence time is increased by a factor of about 2,
underlining the possibility to implement dynamical decoupling also
purely optically. In addition, we find that the spin-coherence
dynamics shows signatures of a deviation from a simple exponential
decay.

The experiments are performed on an ensemble of self-assembled
(In,Ga)As/GaAs QDs, which show a resident hole occupation due to
residual doping by carbon impurities, as demonstrated
earlier.\cite{Varwig2012} The sample contains ten layers of QDs
separated by 100-nm GaAs barriers, each with a dot density of
$10^{10}$\,cm$^{-2}$.  The ground state emission maximum of the
photoluminescence (PL) is at 1.38\,eV with a full width at half
maximum of 20\,meV. The sample is mounted in a cryostat, where it is
cooled to temperature $T=6$\,K and exposed to a magnetic field of $B=1$\,T along
the $x$ direction, normal to the sample growth direction that coincides
with the optical axis $z$ (Voigt geometry).

\begin{figure*}[t]
\includegraphics[width=2.05\columnwidth]{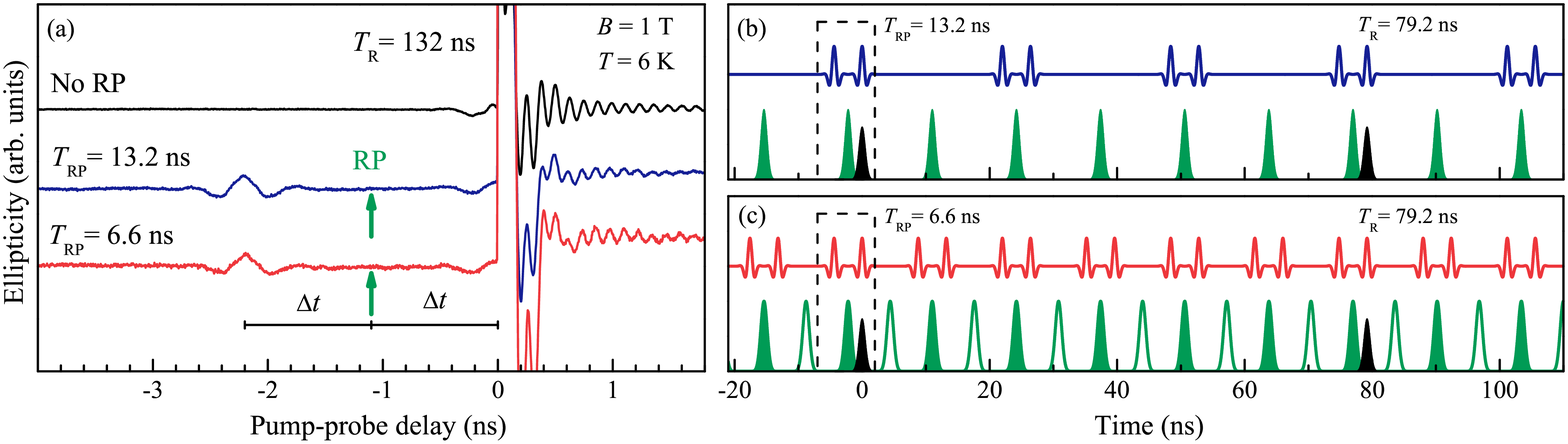}
\caption{(Color online) (a) Time-resolved ellipticity measurements
at $B=1$\,T with pump-probe repetition period $T_{\text{R}} =
132$\,ns at $T=6$\,K. The black, top trace shows the spin
polarization around the pump incidence at zero delay without
applying RPs. The middle, blue trace shows a measurement with an
additional RP train shifted by 1.1\,ns to earlier times (see the
green arrow) relative to the pump train with a RP period
$T_{\text{RP}} = 13.2$\,ns, resulting in a hole spin echo 2.2\,ns
before pump incidence. In the lower, red trace the RP period is
reduced to $T_{\text{RP}} = 6.6$\,ns so that the number of rotations
between two pump pulses is doubled compared to the 13.2\,ns RP period. (b) Scheme of RP application (green) and
echo appearance (blue) for $T_{\text{RP}} = 13.2$\,ns between two pump pulses (black) with $T_{\text{R}} = 79.2$\,ns. (c) Echo appearance (red) for $T_{\text{RP}} = 6.6$\,ns. The additional RP incidences are indicated by the open green pulses.}
\label{fig:spectra}
\end{figure*}

The hole spin dynamics is studied by a degenerate pump-probe setup
employing in the simplest version of the experiment a pulsed
Ti:Sapphire laser with its photon energy tuned to the PL maximum.
The laser emits pulses with a duration of 1\,ps at a repetition rate
of 75.75\,MHz. The pulse period used for the experiment was reduced
by a pulse picker letting every $n$-th pulse pass while blocking all
other pulses in between such that the pulse repetition period
$T_{\text{R}}$ is a multiple of the original period of 13.2\,ns. In
our studies it was varied from 132 to 462\,ns. The laser output is
split into circularly polarized pump and linearly polarized probe
pulses, which can be delayed relative to each other. The pump power
is adjusted to a pulse area of $\Theta=\pi$,\cite{Varwig2013}
the probe power is about ten times weaker.

The train of pump pulses creates a spin polarization along the
optical axis by exciting positively charged excitons (trions). After their
radiative decay, spin-oriented holes are left behind, which
subsequently precess about the magnetic field in the $yz$-plane.
Before the decay, also precession of the optically excited electron spins contributes to the coherent signal.
The spin polarization is monitored by measuring the ellipticity of
the probe pulses acquired by transmission through the sample. The
upper black trace in Fig.~\ref{fig:spectra}(a) shows a
time-resolved ellipticity measurement of the spin polarization with
pump incidence at zero pump-probe delay. After the incidence one
observes dominantly fast precession from photoexcited electrons and superimposed 
slow spin precession from the resident holes, each
corresponding to the characteristic $g$ factors. The observed number
of hole spin oscillations is rather small due to fast dephasing,
arising from the considerable $g$ factor inhomogeneity in the
ensemble. The decay of the electron spin precession is mainly due to
radiative trion recombination. The precession signal shortly before
pump incidence, at negative delays, results from mode-locked hole
spins.\cite{Varwig2012,Eble2012} From the dependence of this mode-locking signal amplitude on the delay between pump pulses the hole spin coherence time can be assessed. The measured value depends sensitively on parameters such as pump power or sample temperature. Due to slight variations in these parameters we typically find variations of the coherence time from a few hundred ns up to a $\mu$s (see also below).

To induce spin echoes we extend the experimental setup by adding 
a further, pulsed Ti:Sapphire laser that is used for optical      
spin rotations.\cite{Greilich2009} This laser is synchronized to
the pump-probe laser with an accuracy of 1\,kHz. The rotation-pulse 
(RP) duration is also 1\,ps. The RP repetition period, however, was 
taken either as emitted from the laser so that the pulses are separated 
by $T_{\text{RP}} = 13.2$\,ns, or the pulse period was reduced to
6.6\,ns by splitting the laser output into two pulse trains, sending
one train along a mechanical delay line of 2\,m length and reuniting
it with the other pulse train.

The goal of this pulse train is to invert the momentary orientations
of the precessing hole spins at a certain moment after initial
orientation. In our case this moment is chosen such that the
macroscopic hole spin coherence is already dephased. To reverse the
dephasing, again excitation of the trion transition by circularly
polarized pulses is applied. To avoid generation of new spin
coherence, but manipulate existing spin coherence only, the area of
these RPs is adjusted to $\Theta=2\pi$. Under these conditions the
system is excited during the pulse action to a positively charged trion and returned back to the
resident hole. After this complete Rabi-flop the hole has acquired a
geometrical phase, that corresponds to a rotation about the optical
axis by an angle of $\pi$ for resonant
excitation.\cite{Economou2007}

As the RP is applied when the spin ensemble is dephased, the hole
spin orientations are arbitrarily distributed in the plane normal to
the magnetic field. In this case the RP action is identical to a
reflection of the spins at the plane spanned by the optical axis and the
magnetic field. Then, the precession after inversion brings the
dephased spins into phase again at time $2 \Delta t$ that is twice
the separation between rotation and pump pulses $\Delta t$. At this
moment, the hole spins are again fully aligned (in case of perfect
RP action on the ensemble, similar to pump pulse application), so that a further RP coming in later can
induce a spin echo again. Since $T_{\text{RP}} \ll T_{\text{R}}$ in
this experiment, the spins are rotated multiple times between two
pump pulses. This leads to a sequence of dephasing and rephasing,
resulting in multiple echoes between two pump pulses.

The RP and echo sequences between two subsequent pumps are shown
schematically in Fig.~\ref{fig:spectra}(b). The upper part
gives the case in which the RP separation is 13.2\,ns, as indicated
by the shaded green pulses equidistantly spaced in time relative to
the black colored pump pulses. For simplicity, the separation
between the pump pulses was assumed to be $T_R=79.2$\,ns, much smaller
than the minimum separation in experiment, for
clarity of discussion. The upper blue curve shows the expected signal
echo sequence as result of the RP application. After pump incidence
at zero delay the first RP hits at 12.1\,ns, leading to echo
formation at 24.2\,ns delay. The next RP comes in 1.1\,ns later at
25.3\,ns delay and manipulates on the spin coherence from the
previous echo, so that it induces the next echo another 1.1\,ns
later at 26.4\,ns. The situation at this echo time is basically
identical to that at the moment of pump pulse application so that
the subsequent RPs repeatedly induce identical echo sequences.

Note that due to the particular periodicity of rotation and pump
pulse application, a peculiar symmetry of the echo appearance in
time arises. With the RP incidence at $\Delta t$, the echo
appearance occurs at $2 \Delta t$, independent of the sign of $\Delta
t$. Furthermore, the echoes occur symmetrically to each RP. This can
both be seen in the lower part of
Fig.~\ref{fig:spectra}(b) showing the case when the RP period is
bisected as indicated by the additional unshaded green pulses.
Consequently the number of echoes is doubled with the additional
echoes appearing just in between the ones in the previous case.

\begin{figure}[t]
\includegraphics[width=\columnwidth]{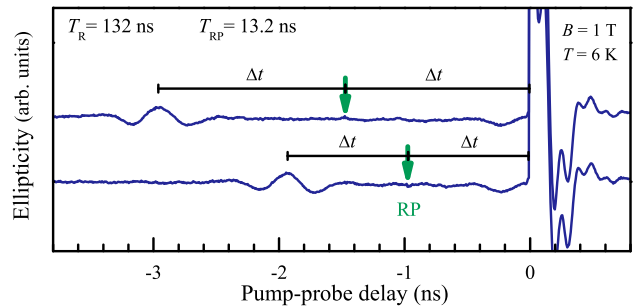}
\caption{(Color online) Spin echo emergence at time $2 \Delta t$ for
different arrival times $\Delta t$ of the RP, $-1.5$\,ns and
$-1$\,ns, as indicated by the green arrows before pump pulse arrival
at zero delay. $B$ = 1\,T at $T$ = 6\,K.} \label{fig:position}
\end{figure}

In the experiments the length of our mechanical delay lines for
adjusting the different pulse trains relative to each other allows
us to cover a delay range of about 6\,ns, which therefore has to be selected
carefully. We have decided for the range from about
$-$4\,ns to 2\,ns. In both protocols with 13.2\,ns and 6.6\,ns pulse
separation, respectively, a RP comes in at $\Delta t = -1.1$\,ns, so
that an echo (induced by a previous RP) should appear at $2 \Delta t = -2.2$\,ns. The
corresponding experiments are shown by the two lower curves in
Fig.~\ref{fig:spectra}(a). The green arrows indicate the moment in
which the RP hits the sample in the two cases with different
$T_{RP}$. In agreement with the expectations, hole spin echoes are
observed at $-2.2$\,ns in the ellipticity measurements. The echoes
have comparable amplitude and show a similar fast dephasing as the
hole spin coherent signal after a pump pulse. The echo character is
confirmed by varying the moment of RP arrival $\Delta t$, as done in
Fig.~\ref{fig:position}. When changing $\Delta t$ from $-$1\,ns to
$-1.5$\,ns, the echo at $2 \Delta t$ shifts correspondingly from $-2$\,ns to $-3$\,ns.

The RP train sequences are similar to the CPMG protocol originally implemented in
NMR studies.\cite{CPMG1958,footnote} This protocol has recently been used to
keep a quantum bit embedded into surrounding baths alive. Simply
speaking, its impact can be understood such that the RPs invert the
spins within time periods that are short compared to the effective
coupling time to the baths so that the spins become decoupled from
them. The periodic RP sequences used in our experiments allow us to
assess the potential of optical dynamic decoupling protocols for extending spin
coherence.

\begin{figure}[t]
\includegraphics[width=\columnwidth]{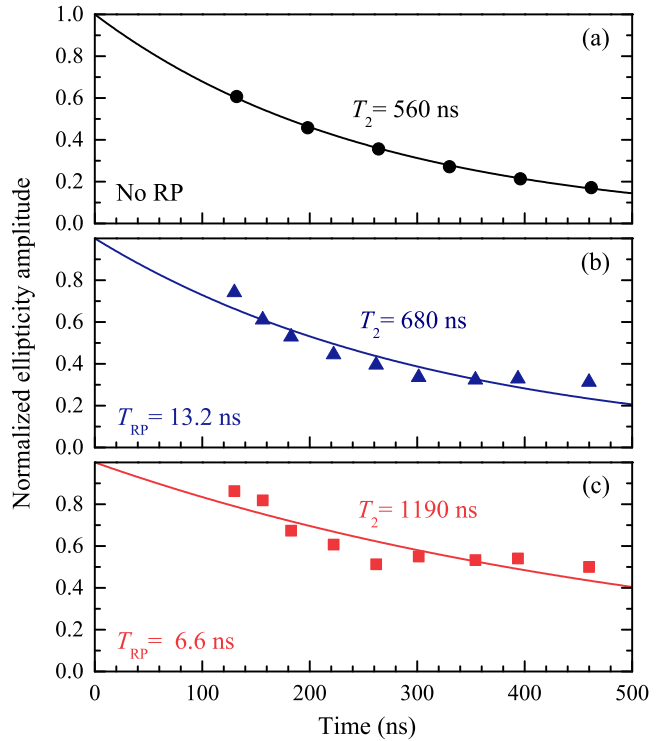}
\caption{(Color online) (a) Normalized ellipticity amplitude of mode-locked hole-spins right before a pump pulse in dependence of the time after previous pump incidence
$t=T_{\text{R}}$ without RP application. (b) Normalized echo amplitudes in dependence of the time after previous pump incidence $t=T_{\text{R}} +2
\Delta t$ with $\Delta t = -1.1$\,ns for a RP separation of $T_{\text{RP}}= 13.2$\,ns. (c) Normalized echo amplitudes for $T_{\text{RP}}= 6.6$\,ns.}
\label{fig:ampl}
\end{figure}

To measure the hole spin coherence time, the period between two pump pulses
$T_{\text{R}}$ is increased using the pulse picker and the amplitude
of the last echo before pump incidence at $-2.2$\,ns delay is
measured as a function of $T_{\text{R}}$. For reference, also the hole-spin mode-locking amplitude right before zero delay as a function of the separation between pump pulses without RP application is shown in Fig.~\ref{fig:ampl}(a). 
These data were recorded under the same experimental conditions (pump power etc), optimized for 
the longest possible coherence time achievable, as in the echo studies. The data are normalized to unity for zero delay.
Within the scanned range of pump separations, the mode-locking amplitude decays due to the loss of coherence.
To assess the underlying coherence time quantitatively we fit the data
for the mode-locking amplitude $A(t=T_{\text{R}})$
by an exponential decay. In detail, we use the fit form that was
elaborated in Ref.~[\onlinecite{Greilich2006}] for mode-locked
spins:
\begin{equation}
A(t) = A_0
\exp\left[-\left(2+\frac{1}{2\sqrt{3}+3}\right)\frac{t}{T_2}\right], \label{eq:exp}
\end{equation}
with the hole-spin coherence time $T_2$. Due to the normalization of
the data, $A_0 = 1$. From this fit we obtain a hole spin coherence time of $T_2=
560$\,ns, in reasonable accord with previous reports.\cite{Varwig2012,Varwig2013} This value serves as reference value for all subsequent experiments involving spin echo signals. 
For that purpose we accurately determine in every experiment the hole-spin mode-locking amplitude and normalize it to this reference value to get rid 
of possible variations in experimental parameters. 

The amplitudes of the last echoes before pump incidence induced by the two sequences of RP application with  $T_{\text{RP}}= 13.2$\,ns and  $T_{\text{RP}}= 6.6$\,ns are plotted as a functions of the time after the last pump incidence in the panels (b) and (c) of Fig.~\ref{fig:ampl}, respectively. This time is given by $t=T_{\text{R}} +2
\Delta t$ with $\Delta t = -1.1$\,ns. Again, we have normalized the two data
sets such that right after pump action and before the RPs can have
an effect, the amplitudes are identical.
 This normalization has no impact on the exponential decay times. Also here one can see the decay of the echo amplitude with increasing pump separation $T_R$, but clearly the drop occurs much
faster when the RP sequence with $T_{\text{RP}} = 13.2$\,ns is
applied compared to the 6.6\,ns case. This indicates that indeed the
coherence of the hole spins is kept alive more efficiently for short
RP separation. We use the same fit form as for the mode-locked signal to assess the coherence time in these experiments. For the case
with 13.2\,ns RP separation we obtain a coherence time of $T_2=680$\,ns from the fit of the experimental data which is only slightly longer than for the case without rotation pulses.
When the separation between RPs is bisected, the coherence time increases by about a factor of
two up to 1190\,ns, more than twice the time without rotation pulses.

This result clearly supports decoupling from surrounding baths by periodic laser pulses. The limitation of the hole spin
coherence at $T$ = 2\,K with a quick drop
for temperatures higher than liquid helium has been tentatively
assigned to the hyperfine interaction with the
nuclei.\cite{Varwig2013} On the other hand, this interaction has
been shown to be about an order of magnitude weaker than for
electrons,\cite{Fischer2008,Eble2009,Fischer2010,Fallahi2010,Chekhovich2011}
even though the observed spin coherence times are very much
comparable for electron and hole spins.\cite{DeGreve2011,Varwig2012}
This underlines that further work needs to be done in this field to
understand the coupling of spins to baths on a microscopic level.
Dynamic decoupling techniques might be helpful in this respect:
While the mode-locking data [Fig.~\ref{fig:ampl}(a)] can be quite well described by
an exponential decay, the data recorded with RP application indicate that the echo
amplitude decrease does no longer follow a pure exponential function
but appears more complicated with an initial somewhat faster drop
followed by a much slower decrease. By fitting the data points for times exceeding 200\,ns, decay times exceeding 10\,$\mu$s are obtained. Further studies are required to understand this behavior in more detail.

In summary, we have demonstrated the potential of all-optical
protocols for decoupling the dynamics of QD-confined carrier spins
from the surrounding baths. While much shorter pulse durations can
be obtained in that way, the main obstacle at the moment is the less
accurate and flexible possibility to tailor optical pulses compared
to the microwave range. However, the huge progress in
pulsed laser technology might open novel perspectives here. This
progress may be used to implement more complex dynamic decoupling
protocols that are optimized for keeping the coherence of a quantum
bit alive.

This work was supported by the Deutsche Forschungsgemeinschaft and
the BMBF program Q.com-H. M.B. acknowledges support from the Russian Ministry of Education and Science (contract No.14.Z50.31.0021).

\bibliographystyle{apsrev}
\bibliography{DD_refs}

\begin{thebibliography}{32}
\bibitem[{foo()}]{footnote}
\bibinfo{note}{In the CPMG protocol the periodic pulse sequences are composed
  not only of a single pulse as in our case, but of pulse triplets by which the
  spins are excited first to a coherent superposition of the spin basis states
  by $\pi/2$-pulses, then inverted by $\pi$-pulses and finally rotated again
  into a basis state by another $\pi/2$-pulse. In our case, the spins are
  precessing about the magnetic field and the rotation pulses are applied when
  the spins are already randomized in the plane normal to the field.}
\end{thebibliography}

\end{document}


\title{Comment on the different versions of the article}







\maketitle


The absolute coherence times depend on several experimental parameters such as crystal purity, sample temperature, laser power, etc. As a result of changes in these parameters we find variations of the hole spin coherence time from about 200 ns up to 1 $\mu$s, measured by the spin mode-locking effect in the quantum dot ensemble (without implementing spin echo protocols).

With increasing complexity of the laser-pulse protocol, the number of possible variation parameters in excitation increases. In the best possible implementation of the used setup that allows also application of multiple periodic pulses for dynamic decoupling we achieved coherence times of about 600 ns. If these optimized conditions are not met, the coherence times can be considerably shorter as in the first version of the manuscript where mode-locking quickly disappeared with increasing pump separation.

Irrespective of the concrete conditions, the dynamic decoupling protocol with a separation of 13.2 ns between the pulses for spin inversion rendered coherence times comparable to or slightly longer than the coherence time without such protocols. Furthermore, the coherence times increased by a factor of about 2 when the protocol with 6.6 ns inversion pulse separation was applied.